# Metastable diamagnetism in $Sm_{0.1}Ca_{0.84}Sr_{0.06}MnO_3$ manganite

V. Markovich,[1] I. Fita,[2, 3] R. Puzniak,[2] C. Martin,[4] K. Kikoin,[1] A. Wisniewski,[2] S. Hebert,[4] A. Maignan,[4] and G. Gorodetsky[1]

[1]*Department of Physics, Ben-Gurion University of the Negev, 84105 Beer-Sheva, Israel*

[2]*Institute of Physics, Polish Academy of Sciences, 02-668 Warsaw, Poland*

[3]*Donetsk Institute for Physics and Technology, National Academy of Sciences, 83114 Donetsk, Ukraine*

[4]*Laboratoire CRISMAT, UMR 6508, ISMRA, 14050 Caen Cedex, France*

## Abstract

Magnetic properties of polycrystalline $Sm_{0.1}Ca_{0.84}Sr_{0.06}MnO_3$ in pristine and metastable states have been investigated in wide range of temperatures and magnetic fields. It was found that below Curie temperature $T_C \approx 105$ K the pristine state exhibits phase separation comprising ferromagnetic and antiferromagnetic phases. The metastable states with reduced magnetization were obtained by successive number of quick coolings of the sample placed in container with kerosene-oil mixture. By an increasing number of quick coolings (> 6) the long time relaxation appeared at 10 K and the magnetization reversed its sign and became strongly negative in wide temperature range, even under an applied magnetic field of 15 kOe. The observed field and temperature dependences of the magnetization in this state are reversed in comparison with the ordinary ferromagnetic ones. Above $T_C$, the observed diamagnetic susceptibility of the reversed magnetization state at $T = 120$ K is $\chi_{dia} \approx - 0.9 \times 10^{-4}$ emu $g^{-1}$ $Oe^{-1}$. Only after some storage time at room temperature, the abnormal magnetic state is erasable. It is suggested that the negative magnetization observed results from a specific coupling of the nano/micro-size ferromagnetic regions with a surrounding diamagnetic matrix formed, in a puzzled way, by the repeating training (quick cooling) cycles.

The group of electron-doped rare earth manganites of the general formula $Ln_xCa_{1-x}MnO_3$ (Ln = rare earth, $x < 0.5$) has attracted considerable attention because of their outstanding magnetic and electric properties. [1-5] The electron-doped manganites do not exhibit a pure ferromagnetic (FM) ground state at any composition. Manganites with $0 < x < \sim 0.02$ have a $G$-type antiferromagnetic (AFM) spin structure. With increasing $x$, local FM regions appear in the $G$-type AFM matrix. For $0.07 < x < 0.16$ and at low temperatures the magnetic spin structure comprises $G$-type AFM, $C$-type AFM, and local FM regions, which enhance around $x \approx 0.1$ at the expense of the AFM phase.[2-5]

The properties of phase-separated manganites are governed to a large extent by the averaged radii of the A-site cations, $r_A$, of $ABO_3$ perovskite structure and by the disorder of their distribution.[1,2] The quenched disorder in a solid solution of rare earth and alkaline earth (AE) elements can be evaluated by the cation disorder parameter $\sigma^2 = \sum_i (x_i r_i^2 - r_A^2)$, where $x_i$ describe fractional occupancies of A-site cation position and $r_i$ are the effective ionic radii of A-site cations, i.e, of Ln and of Ca. [6] In this paper, we present the results of the measurements of magnetic properties of $Sm_{0.1}Ca_{0.9-y}Sr_yMnO_3$ with low Sr doping ($y = 0.02, 0.06, 0.1$). The increase of Sr content results in the increase of $r_A$ (from $r_A = 1.332$ Å for $y = 0.02$ to $r_A = 1.34$ Å for $y = 0.1$) and the $\sigma^2$ changes accordingly from $\sigma^2 = 0.0011$ for $y = 0.02$ to $\sigma^2 = 0.002$ for $y = 0.1$. Our attention is mostly focused on $y = 0.06$ sample for which an unusual metastable diamagnetic (DIA) behavior under a special thermal treatment is described below.

The magnetic measurements were performed in the temperature range 4.2 - 250 K and magnetic fields up to 16 kOe, using PAR (Model 4500) vibrating sample magnetometer (VSM). All of our measurements were carried out on cylinder-shape samples having a diameter of 1 mm and height of 4.0 mm. [7] Measurements with VSM were carried out on samples placed in miniature CuBe high-pressure container, served as a sample holder, filled with kerosene-mineral oil mixture. Additional measurements of ac susceptibity of $y = 0.06$ sample were carried out using a Quantum Design PPMS.

The measurements of the temperature dependences of the magnetization of $Sm_{0.1}Ca_{0.84}Sr_{0.06}MnO_3$ in as-synthesised state were performed according to the following procedure: the samples were cooled at zero magnetic field to $T = 5$ K and the magnetization was measured upon heating (ZFC curve) and immediately thereafter upon cooling (FC curve) under an applied magnetic field of $H = 100$ Oe.

In similarity with $Sm_{0.1}Ca_{0.9}MnO_3$ sample both Curie and Neel temperatures of $Sm_{0.1}Ca_{0.84}Sr_{0.06}MnO_3$ were equal to about 105 K, see Fig. 1 (a). An inset in Fig. 1(a) presents magnetization curves of $Sm_{0.1}Ca_{0.84}Sr_{0.06}MnO_3$ vs. magnetic field at 10 K and 80 K characterized by a negligible coercive field $H_C$. The spontaneous magnetization $M_0$ obtained by a linear extrapolation of the high field magnetization to $H = 0$ is attributed to the FM phase. Interestingly, the spontaneous weak magnetization ($M_0 = 0.77$ $\mu_B$/f.u. at 10 K), is comparable with the value obtained for $Sm_{0.1}Ca_{0.9}MnO_3$, indicating that both compounds are in similar magnetic state with close ratio of FM/AFM phases. The results obtained for the real part of the ac susceptibility $\chi^{'}$ vs. temperature, at different frequencies are presented in Fig. 1(b). A sharp peak in $\chi'$ at around 103 K is plausibly associated with the critical softening of the FM system near $T_C$. The characteristic splitting between $M_{FCC}$ and $M_{ZFC}$ below $T_C$ (Fig. 1(a)) and distinct frequency dependence of $\chi^{'}$ (Fig. 1(b)) are indicatives of a formation of basically cluster-glass like state characteristic for low electron doped phase separated systems. [1,2,4,5,8]

Further measurements of magnetization of $Sm_{0.1}Ca_{0.84}Sr_{0.06}MnO_3$ were performed on samples placed in a pressure cell filled with the pressure transmitting medium (a mixture of mineral oil and kerosene). The cooling down was performed with very quick cooling (QC) (we estimated that speed of cooling was as high as up to 80 K/min.), resulting in a long time relaxation and in a reduced magnetization, see Fig. 2 and Fig. 3(a). It was found that the first QC thermal cycles do not influence the magnetic properties of the sample: the state with reduced (and eventually reversed magnetization) arises only after 5 - 6 QC cycles (see Fig. 2). States with reduced magnetization have some kind of write-in memory, which may be erased by heating up of the sample to room temperature (RT) and keeping at such temperature for a short time, say 20 - 40 minutes. Sequential additional QC cycles resulted in negative magnetization at low temperatures. Maximal negative value of magnetization was obtained after several (> 6) QC cycles, while keeping the sample at RT during few hours resulted in the formation of a new pristine-like state. It appears that metastable state obtained by applying QC procedure is relatively stable at temperatures 5 – 200 K; above 200 K a switching to other metastable state or to pristine-like state may occur (see an example of such switching at $T \sim 220$ K in Fig. 3(b) and in extended scale in inset of Fig. 3(b)). It is worth noting that even after heating up to 240 K and subsequent cooling (not QC) to low temperatures, diamagnetism may survive, see Fig. 3(b). We have checked the possibility to induce the metastability and negative magnetization in $Sm_{0.1}Ca_{0.9-y}Sr_yMnO_3$ with different Sr doping. We tested various samples but this peculiar diamagnetism was observed only for $y = 0.06$, while for the samples with close composition ($y = 0.02$ and $y = 0.1$) only metastable states with strongly reduced magnetization were obtained.

Figure 4 shows the $M(H)$ curves of $Sm_{0.1}Ca_{0.84}Sr_{0.06}MnO_3$ at various temperatures observed after several QC cycles. The following features are noticeable: (i) a linear diamagnetic response is observed for $T > 110$ K up to magnetic field of 15 kOe. One finds the dc susceptibility $\chi_{dia} \approx -0.9 \times 10^{-4}$ emu $g^{-1}$ $Oe^{-1}$ at $T = 120$ K. At temperatures below 100 K there is an addit011onal strong low-field DIA contribution to susceptibility and the total DIA susceptibility of the sample becomes strongly field dependent.

The occurrence of magnetization reversal and DIA response in perovskite structures is not without precedent. The reversal of magnetization was observed in perovskite vanadates $LaVO_3$,[9] $LuVO_3$,[9] $YVO_3$,[9,10] and also in the metallic glass PrAlNiCuFe.[11] Though the above vanadates are isostructural, the DIA low temperature states observed are found to appear due to different underlying mechanisms. It was suggested that for $LaVO_3$ "anomalous diamagnetism" occurs as a result of the first order Jahn-Teller transition at 138 K below $T_N \approx 142$ K that enhances $V^{3+}$ ion orbital angular momentum and the spin-orbit coupling, leading to reversal of the Dzyaloshinski vector.[9] For YVO$_3$ multiple temperature-induced magnetization reversals at 77 K and 95 K ($T_N \approx$ 114 K) in small magnetic fields were attributed to two coupled spin-canting mechanisms that oppose one another, namely an antisymmetric Dzyaloshinski–Moriya exchange and single-ion anisotropy.[9,10] Reversal of magnetization and DIA response was observed previously also in different manganites, such as (Dy,Ca)MnO$_3$,[12,13] Gd$_{0.67}$Ca$_{0.33}$MnO$_3$,[14] La$_{1-x}$Gd$_x$MnO$_3$,[15] NdMnO$_{3+\delta}$,[16] Nd$_{1-x}$Ca$_x$MnO$_3$,[17] and (Nd,Ca)(Mn,Cr)O$_3$.[18] The reversal of magnetization in these compounds was interpreted as a manifestation of a ferrimagnetic-like behavior due to the interplay of two magnetic antiferromagnetically coupled sublattices (rare–earth sublattice and Mn sublattice). If magnetizations of two sublattices have different temperature dependences, and at decreasing temperature the magnetization of the sublattice aligned antiparallel with magnetic field (rare–earth sublattice) grows more rapidly than that aligned with field, the net moment may point opposite to the applied magnetic field in a ferrimagnetic-like state below the compensation temperature ($T_{comp}$).[12-16] In the case of metallic glass of PrAlNiCuFe[11] it was suggested that a specific coupling between the FM nanoparticles and the amorphous matrix might account for a giant diamagnetic response.

It should be noted that principal distinction between the DIA response of the above samples and that of Sm$_{0.1}$Ca$_{0.84}$Sr$_{0.06}$MnO$_3$ consists in the fact that the metastable DIA response of the later one occurs only under very critical conditions, namely, the quick cooling (~ 60-80 K/min.) of the sample while it is being placed in pressure cell filled with pressure transmitted media (50/50 mixture of kerosene and mineral oil). The absence of one of these conditions is crucial: for example, QC without a pressure transmitted media does not lead to appearance of diamagnetic effect. Besides, the preparation conditions are probably crucial for observation of anomalous diamagnetism. The striking features of diamagnetism in $y = 0.06$ sample are also the relaxation of magnetization at low temperatures and wide temperature range of existing diamagnetic state.

One may assume that the quick solidification of the pressure transmitted media between the sample and the inner wall of the pressure cell produces inhomogeneous strains in the sample, which account for metastable states and DIA response. Generally, the reduction of pressure during cooling to 4.2 K in a clamped low-temperature piston-cylinder pressure cell is about 20-30% (independently on the construction of the cell). [19] In our studies the pressure after clamping the pressure cell at room temperature was ~ $0.04 - 0.05$ GPa, and therefore the resulting pressure at low temperature is close to zero. [19,20] Nevertheless, we believe that in our experiments the diamagnetic state occurs due to anisotropic stresses which may arise upon quick solidification of the kerosene-oil mixture. These stresses in low-temperature cells are the combined result of many factors: unequal and non-uniform contraction of the cell and of the pressure transmitted media, the rate of the cooling, and the placement and the dimension of the sample. [20] The observed switching from one metastable state to another above $T_C$, supports this view, see Fig. 3(b). The freezing/melting of the transmitted media might be the cause for the switching between metastable states. It is known that melting point ($T_{melt}$) of kerosene is about 235 K. [21]

A few observations of large diamagnetism in metastable states have been reported previously for CuCl [22-24] and CdS [25] and ball-milled $Sr_{0.6}Ca_{0.4}CuO_2$. [26] Diamagnetism in CuCl and CdS was interpreted as a possible manifestation of an excitonic mechanism of superconductivity. [22,24,25] Chu *et al.* [23] claimed that results observed for CuCl may be explained by superconductivity at the Cu-CuCl interface due to the disproportionation of $2CuCl \rightarrow Cu + CuCl_2$. As a rule DIA anomalies were observed only when the CuCl samples were rapidly cooled or warmed. [22-24] Brown *et al.* [25] observed large diamagnetism approaching 100% flux exclusion in CdS at 77 K, after sudden release of pressure with the rate greater than 100 GPa/sec. It appears that in similarity with the results for $Sm_{0.1}Ca_{0.84}Sr_{0.06}MnO_3$ a strong diamagnetism in CuCl and CdS was observed only in highly inhomogeneous state. As a plausible alternative to the models assuming excitonic mechanism of superconductivity [24,25,27] in CuCl and CdS, the electron localization mechanism, resulting in strong diamagnetism was also suggested. [28,29] Chaban [28] has formulated the description of DIA susceptibility by the expression: $\chi = -\tilde{n}e^2 <p^2>/6mc^2$, where $\tilde{n}$ is the number of electron per $cm^3$, $<p^2>$ is the mean square radius of the electron orbit, and $m$ is the electron mass. This formula applied for polycrystalline CuCl with grain radius of the order of 100 nm gave $\chi = -10^{-2}$ in reasonable agreement with experimental results.[22-24,28] Similar explanation was proposed for large DIA susceptibility in ball-milled $Sr_{0.6}Ca_{0.4}CuO_2$.[26] Hernando *et al*.[26] suggested that the large size electron orbits of the nanostructures originated by the ball milling process account for the DIA response. The microcrystalline size of the ball-milled samples was estimated to be around 15 nm. Scanning electron microscope (SEM) micrographs show that the typical microstructure size of our sample of $Sm_{0.1}Ca_{0.84}Sr_{0.06}MnO_3$ is of about 10 μm at RT (Fig. 5), too large for being a reasonable diameter of electron orbitals. The crystallites of nano dimension were not resolved by the SEM measurements.

Fisher *et al*.[30] argued recently that fast cooling results in a large number of defects and quenched by fast cooling disorder leads to formation of inhomogeneous metastable states in $(Sm,Sr)MnO_3$. In this case the reasonable radius of large size orbits may be rather related to a size of extended defects and exceeds tens of nanometers. It should be emphasized that the existence of large-size orbits must not be accompanied by an increased macroscopic conductivity as the latter is probably governed by the high resistivity of grain boundaries and regions between grains. The calculations of Onishchenko[29] for a cubic crystal with narrow valence and conduction bands have revealed that the orbital motion of band electrons in a macroscopically inhomogeneous crystal can produce an anomalously strong diamagnetism, which may not be associated with electron-electron interactions.

Basing on experimental data and above discussion, one may suggest that in highly inhomogeneous state like that occuring in $Sm_{0.1}Ca_{0.84}Sr_{0.06}MnO_3$ the observed strong diamagnetism may be related to metastable nanocrystalline configuration resulting in localized wave function of electrons with large size electron orbits. Indeed, when examining the experimental data for the field dependent negative magnetization shown in Fig. 4, we see that the sample behaves in the metastable state as a system of FM clusters in a DIA matrix. Below FM transition with $T_C \approx 100$ K (see Fig. 1) the magnetization curves exhibit a high negative permeability at low fields and a DIA susceptibility at higher temperatures. In some sense the magnetization curves look like "mirror" replica of the characteristic $M(H)$ dependence of the ordinary ferromagnetic curves.

In order to verify the nature of the negative magnetization we have compared the temperature dependences of the magnetization at $H = 100$ Oe for the pristine and for the DIA states, see Fig. 6. It was found that the absolute values of $M(T)$ in the pristine state and normalized values for the DIA state almost coincide, see Fig. 6. This seems to be a clear evidence that diamagnetic response is associated with magnetization of FM clusters.

The dislocation network formed in strained polycrystalline sample in a contact with frozen kerosene-oil mixture is a plausible medium where electrons with large scale orbits may be confined. Such electrons localized in orbits with radius of tens nanometers can give rise to the linear diamagnetic response at temperatures $T_C < T < 240$ K, see Fig. 4. The negative magnetization observed at $T < 110$ K may result from the phase separation in which the FM clusters couple to the diamagnetic matrix. The magnetization of the FM clusters at $T < T_C$ is almost saturated in magnetic field $H > 5$ kOe, and therefore the negative value of $dM/dH$ at $H > 5$ kOe is seemingly attributed to the DIA matrix. It appears that the slopes $dM/dH$ at 120 K (- 0.9 $\times 10^{-4}$ emu $g^{-1}$ $Oe^{-1}$) and at 10 K for $H > 5$ kOe (- 0.7 $\times 10^{-4}$ emu $g^{-1}$ $Oe^{-1}$) (Fig. 4) have similar values in support of the above model.

The anomalous diamagnetism arises in the highly inhomogeneous strained sample due to QC cyclings. It is reasonable to suggest that a dislocation network having a very high density $n \sim 10^{14}$ - $10^{15}$ $m^{-2}$ may be created in the inhomogeneously strained sample. The average distance between dislocation in a network of density $n$ is of the order of $1/\sqrt{n}$, i.e. 100 nm and 32 nm for $n$ equal to $10^{14}$ and $10^{15}$ $m^{-2}$, respectively. [31] Then the appearance of localized electron orbits with radius of tens nanometers, confined within dislocation network, seems a realistic assumption. The presence of some kind of diamagnetic memory, observed for 20 - 40 min. at RT may be associated with the dislocation relaxation via thermal activation. [31,32] Storing the sample at RT for one hour or more results in a rejuvenation of the sample from high density dislocation networks and in an erasing of the diamagnetic memory.

The stresses in our experiments remain probably relatively low and they cannot rather exceed the values corresponding to the reduction of pressure during cooling to 4.2 K in a clamped low-temperature piston-cylinder pressure cell due to the different thermal expansions of the pressure medium and cell components (~0.1 – 0.2 GPa).[20] On the other hand, induced by QC procedure stresses should undertake the high enough dislocation velocity for the creation of the dislocation network. The stress dependence of dislocation velocity varies significantly from one materials to another, see Ref. 31. Studies on face-centered cubic and hexagonal close-packed crystals have shown that the dislocation velocity $v$ varies as $v = A\tau^m$, where $A$ is a material constant, $\tau$ is the applied shear stress, and m is approximately 1 at 300 K in pure crystals, and increases to value from the range 2 - 5 with alloying and to 4 -12 at 77 K.[31] The strong rise in $v$ with decreasing temperature is attributed to a decrease of the dominant damping forces, it arises from the scattering of phonons. Experimental results for various materials have shown[31] that dislocation velocity increases significantly just in range of small shear stress - 0.0001 – 0.1 GPa. For example, in NaCl at room temperature the dislocation velocity increases from $10^{-10}$ ms$^{-1}$ to $10^{-1}$ ms$^{-1}$ at applied shear stress increasing from about 0.0003 GPa to 0.003 GPa. It appears that just at relatively low shear stresses the dislocation velocity increases significantly and dislocation network may be created.

It should be noted that negative magnetization was not observed after numerous (> 10) QCs at a relatively modest applied hydrostatic pressure (~ 0.5 GPa). This could be understood within the dislocation network scenario taking into account that the mobility of dislocation decreases under high pressure. It was found[33,34] that under high hydrostatic pressure ~ 1 GPa the mobility of dislocations is decreased by few orders of magnitude. Experimental and theoretical results[33] indicated that the pressure effect increases with the growing concentration of dislocations due to interactions between dislocations and local barriers. The magnetic response to the external field at $T < T_C$ is the sum of DIA and FM contributions, and the relaxation in the metastable state (Fig.

3(a)) is a complex process, which cannot be characterized by a single relaxation time. Both the dislocation network and FM clusters may contribute to the long time relaxation of the magnetization.

It is known that the dislocation networks can change the electronic properties of metallic compounds. As an example of the influence of dislocations on the macroscopic properties of materials one may mention the possibility of formation of non-uniform localized superconducting state in the vicinity of isolated dislocation or dislocation network discussed by Nabutovsky and Shapiro. [35]

In order to provide an insight into the nature of the negative magnetization in $Sm_{0.1}Ca_{0.84}Sr_{0.06}MnO_3$ an alternative magnetostrictive effect was considered. The magnetostrictive approach (Ref. 36) was used for a thin film of $La_{2/3}Ca_{1/3}MnO_{3-\delta}$, it suggests that the cause for negative change in the magnetization ($dM/dH < 0$) may be attributed to the interfacial strain. It was proposed that the internal strain tends to increase the magnetic moment, while the magnetic field works opposite to the stress. In our case of bulk $Sm_{0.1}Ca_{0.84}Sr_{0.06}MnO_3$ sample the absolute value of the magnetization always increases upon applying a magnetic field in contradiction to the model proposed by Poddar *et al.* [36]

In conclusion, we have presented new results concerning the metastablity and diamagnetic behavior of polycrystalline $Sm_{0.1}Ca_{0.84}Sr_{0.06}MnO_3$, observed after several cycles of quick cooling. It has been established that $Sm_{0.1}Ca_{0.84}Sr_{0.06}MnO_3$ exhibits a magnetic phase separation below $T_C \approx 105$ K, involving ferromagnetic and antiferromagnetic phases. We have shown that the sample placed in a capsule with kerosene-oil mixture, exhibits a strong negative magnetization as a result of a number of quick coolings. Surprisingly, the state of negative magnetization is stable for a long time and it is erasable only at temperature above 240 K after keeping the sample at this temperature for about one hour. It is also found that the negative magnetization state which arises as a result of sample training is characterized by a long relaxation of several hours, at 10 K. In a view of the existing models for negative magnetization it seems that localized electron orbits with radius of tens nanometers, confined within dislocation network, may give rise to the diamagnetic effects observed at temperatures $T_C \approx 105$ K $< T < 240$ K. At $T < T_C$ the magnetic response to the external magnetic field is controlled mainly by the coupling of FM clusters with the diamagnetic matrix. It is suggested that the anomalous negative magnetization results from a specific coupling of the nano/micro-size ferromagnetic regions with the surrounding diamagnetic matrix formed in a perplexed way by the numerous cycles of quick cooling. Further investigations of the effect of microstructure and of the variation in chemical composition may provide a better insight into the underlying physics.

The authors thank Prof. D. I. Khomskii for useful discussions and Prof. T. A. Tyson for valuable remarks. We would like to thank Dr. D. Mogilyansky for his help with SEM imaging and Dr. D. Grebille for x –ray characterization of the samples. This work was supported in part by the Polish State Committee for Scientific Research under a research project no. 1 P03B 123 30.

**FIGURE CAPTION**

Fig. 1. (color online) (a) Measurements of zero field cooled magnetization ($M_{ZFC}$) followed by magnetization measurements upon cooling at the same field ($M_{FCC}$) of $Sm_{0.1}Ca_{0.84}Sr_{0.06}MnO_3$ in pristine state. Inset shows magnetization curves $M$ vs. $H$ at 10 K and 80 K; (b) Temperature dependence of ac-susceptibility of $Sm_{0.1}Ca_{0.84}Sr_{0.06}MnO_3$ in pristine state at $H = 10$ Oe for various frequencies.

Fig. 2. (color online) The time variation of magnetization of $Sm_{0.1}Ca_{0.84}Sr_{0.06}MnO_3$ at subsequent quick coolings.

Fig. 3. (color online) (a) An example of characteristic relaxation of magnetization of $Sm_{0.1}Ca_{0.84}Sr_{0.06}MnO_3$ after quick cooling to 10 K in zero field; (b) Temperature dependence of magnetization of $Sm_{0.1}Ca_{0.84}Sr_{0.06}MnO_3$ in two independent realizations of metastable diamagnetic state. Inset shows in the extended scale the switching from one metastable state to another.

Fig. 4. (color online) Magnetization curves $M$ vs. $H$ of $Sm_{0.1}Ca_{0.84}Sr_{0.06}MnO_3$ in metastable diamagnetic state at various temperatures.

Fig. 5. SEM micrograph of $Sm_{0.1}Ca_{0.84}Sr_{0.06}MnO_3$ sample.

Fig. 6 (color online) Measurements $M_{FC}$ of $Sm_{0.1}Ca_{0.84}Sr_{0.06}MnO_3$ in the pristine and in the DIA states performed in magnetic field $H = 100$ Oe.

Corresponding author:
Vladimir Markovich.
Department of Physics, Ben-Gurion University of the Negev,
P.O. Box 653, 84105 Beer-Sheva, Israel.
Telephone: +972-8-6472456.
Fax: +972-8-6472903.
E-mail: markoviv@bgumail.bgu.ac.il

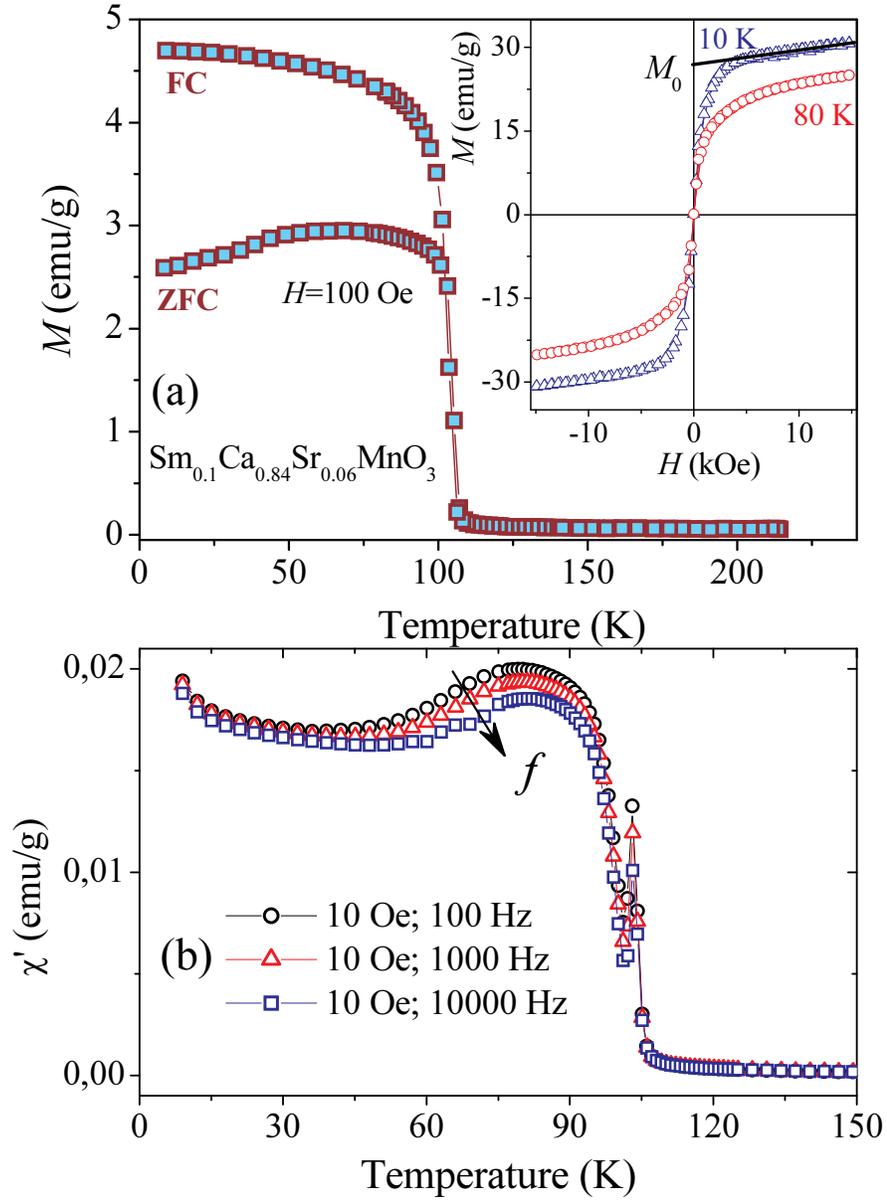

Fig. 1

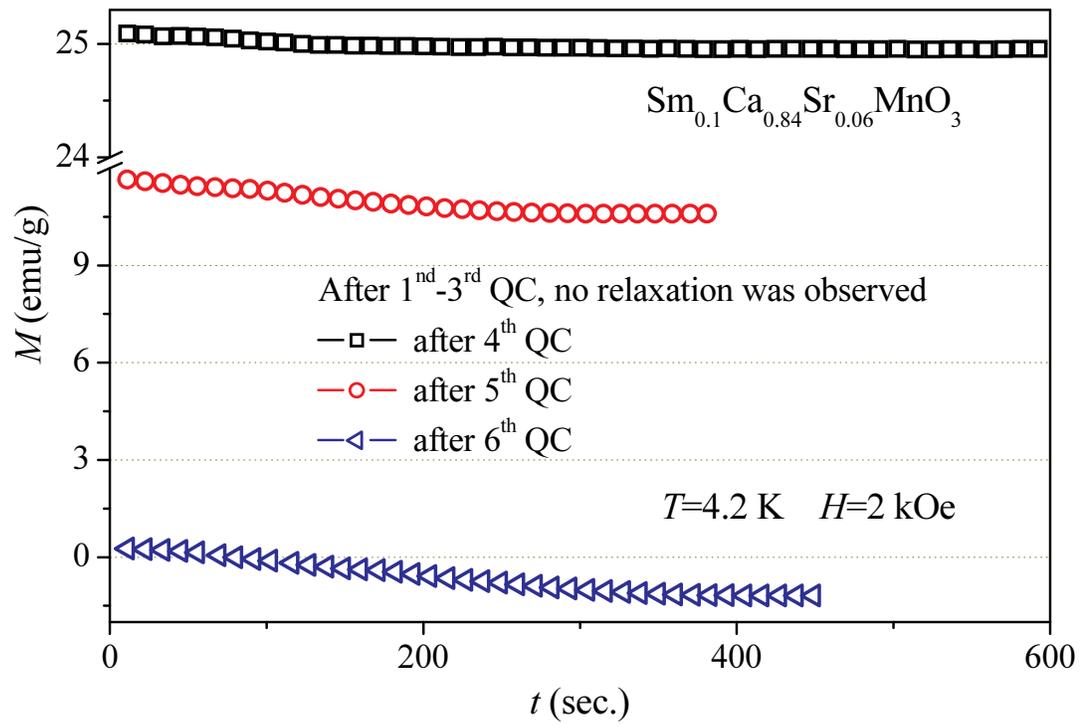

Fig. 2

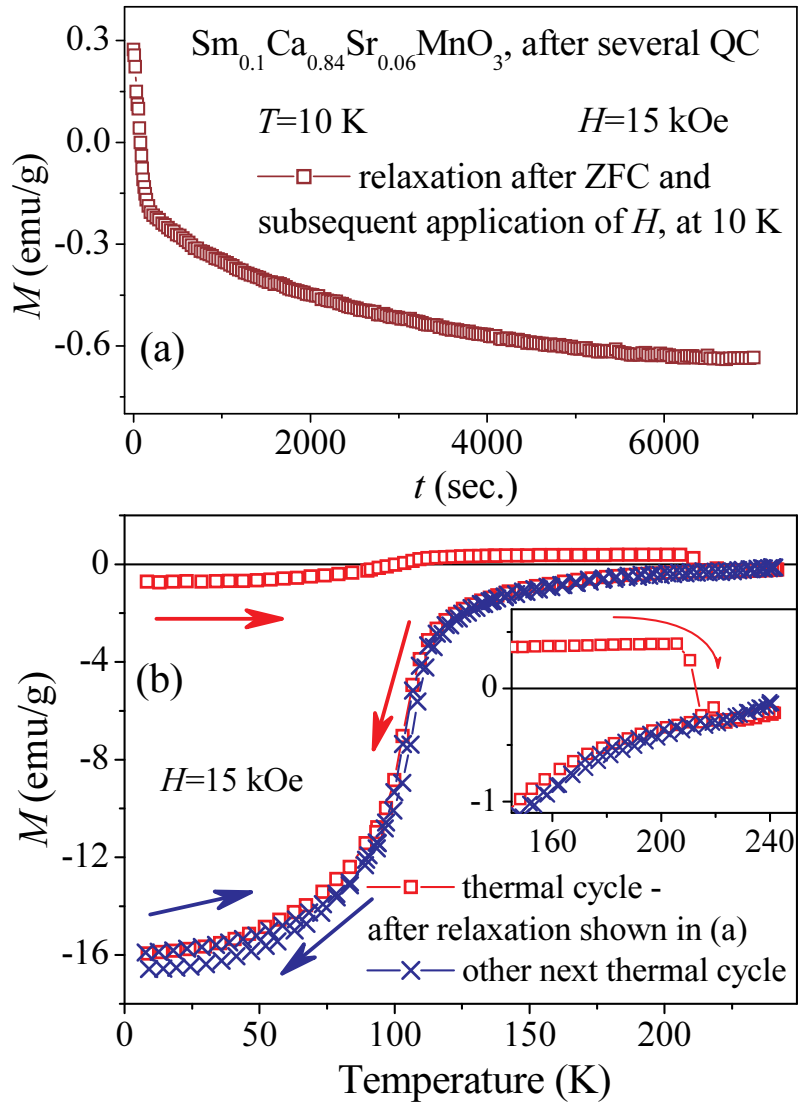

Fig. 3

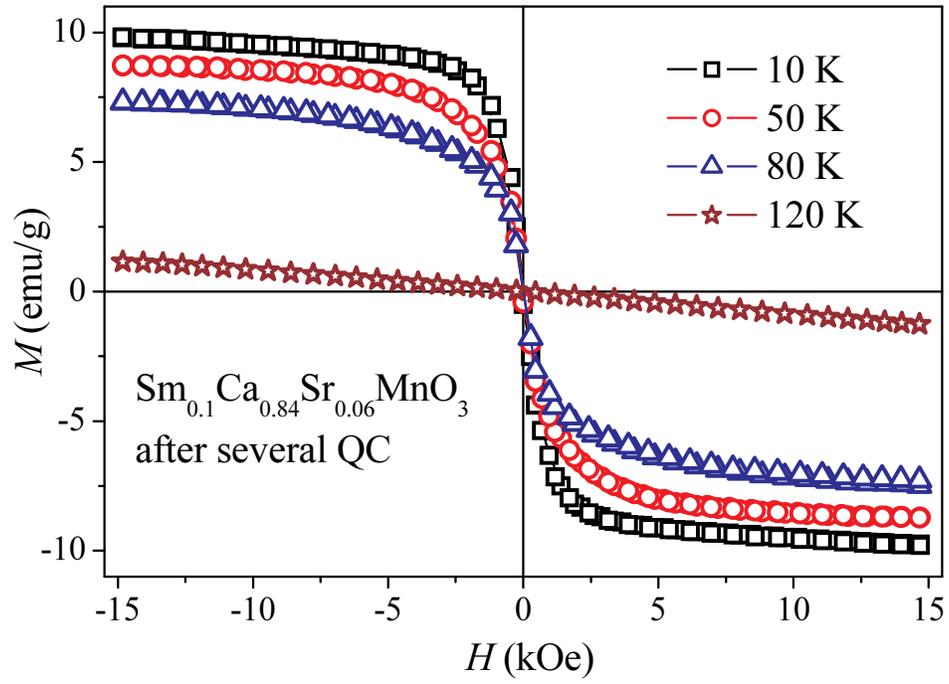

Fig. 4

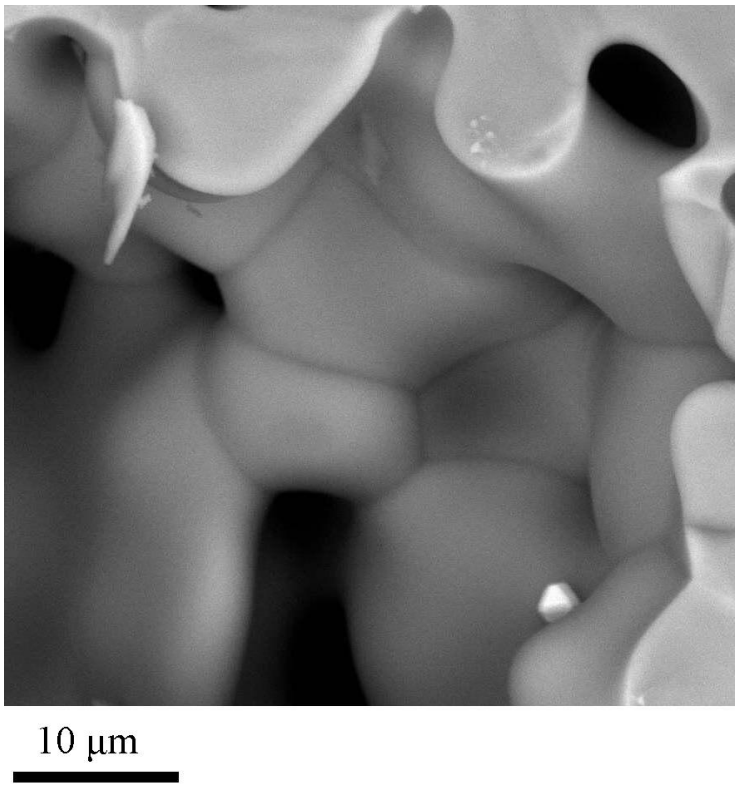

10 μm

Fig. 5

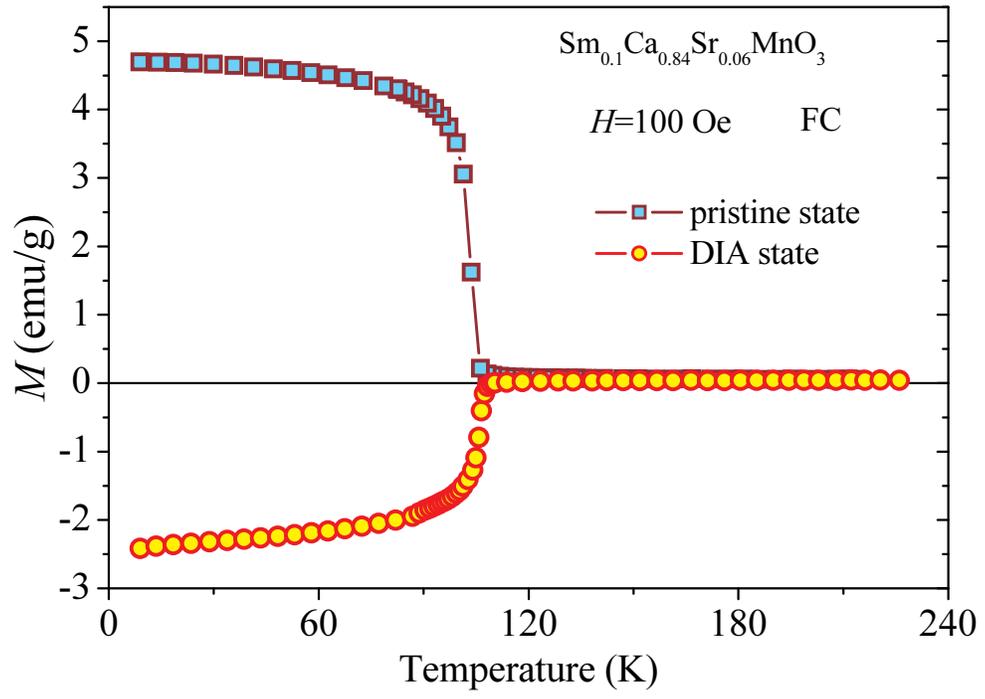

Fig. 6